\documentclass[
    ,final            
  ]
  {aipproc}

\layoutstyle{6x9}

\usepackage{bm}

\begin{document}

\title{Fusion reaction of halo nuclei:\\
A real-time wave-packet method for three-body tunneling dynamics}

\classification{25.60.Pj, 25.70.Mn, 24.10.-i}
\keywords      {Fusion reaction; Projectile and target fragmentation; Nuclear reaction model}

\author{Takashi~Nakatsukasa}{
  address={Institute for Physics and Center for Computational Sciences,
  University of Tsukuba, Tsukuba 305-8571, Japan}
}
\author{Makoto~Ito}{
  address={Institute for Physics, University of Tsukuba, Tsukuba 305-8571, Japan}
  ,altaddress={RIKEN, Hirosawa 2-1, Wako, Saitama 351-0198, Japan} 
}
\author{Kazuhiro~Yabana}{
  address={Institute for Physics and Center for Computational Sciences,
  University of Tsukuba, Tsukuba 305-8571, Japan}
}
\author{Manabu~Ueda}{
  address={Akita National College of Technology, Akita 011-8511, Japan}
}

\begin{abstract}
We investigate fusion cross section of a nucleus with a valence neutron,
using the time-dependent wave-packet method.
For a stable projectile, in which the valence neutron is tightly bound 
($\epsilon_n < -3$ MeV),
the neutron could enhance the fusion probability when
the matching condition of orbital energies are satisfied.
In contrast,
for a halo nucleus, in which the binding energy of the neutron is
very small ($\epsilon_n>-1$ MeV),
the fusion probability is hindered by the presence of the weakly bound neutron.
\end{abstract}

\maketitle


\section{Introduction}

The fusion cross section around the Coulomb barrier energy has been
investigated for a long time \cite{Dag98}.
The cross sections sometimes significantly differ from nucleus to nucleus
depending on properties of nuclear internal structures.
Intrinsic excitations of a projectile and a target during the collision
affect the fusion cross section.
For weakly bound nuclei, excitations to the continuum,
that is breakup as well as transfer process, are expected to play
a very important role.
Since such low-energy reaction cannot be described properly by
(semi-)classical theories,
the full quantum-mechanical reaction theory is necessary.
We have developed a method using the time-dependent wave-packet (TDWP) for
description of the fully quantum reaction dynamics of
a few-body systems \cite{Yab97,YUN03-P,NYIKU04-P,YIKUN04-P,IYNU06}.
An advantage of the TDWP method is given by an intuitive visualization of
dynamics of the quantum reaction.
In practical aspects, the time evolution of the wave packet does not
require complicated scattering boundary conditions.

There have been contradictory predictions on the fusion
cross section of a halo nucleus at sub-barrier energies,
even among calculations using similar models.
We have been studying the fusion probability of halo $^{11}$Be on
$^{208}$Pb \cite{YUN03-P,NYIKU04-P,YIKUN04-P}.
The total fusion cross section (CF+ICF) of $^{11}$Be was predicted to be
smaller than that of $^{10}$Be around the Coulomb barrier energies.
Namely, the presence of a halo neutron leads to a suppression of the
fusion probability.
On the contrary,
the coupled discretized continuum channels (CDCC) calculations,
employing a three-body model very similar to ours,
show a strong enhancement of the total fusion cross section
at sub-barrier energies
for $^{11}$Be on $^{208}$Pb \cite{Hag00,Dia02}.
We have identified that this discrepancy comes from
different truncation for relative angular momentum, $\ell$,
between the neutron and the $^{10}$Be core \cite{YUN03-P,IYNU06}.
Our calculation includes the partial waves up to $\ell=70\ \hbar$,
while the CDCC does up to $\ell=4\ \hbar$ at most.
Actually, when we truncate the partial waves in the same way as
the CDCC calculations do ($\ell\leq 4\ \hbar$), we also
produce the strong sub-barrier fusion enhancement.
However, this enhancement completely disappears as we include higher
partial waves in the calculation.

Importance of the higher partial waves seems to suggest
that the neutron breakup and transfer from $^{11}$Be take place in
an oriented manner.
In fact, we have found that the halo neutron is emitted toward the
target, then some part of the neutron wave packet is caught by the target
potential (``transfer'') and the rest escapes to the free space (``breakup'').
This leads to a simple spectator picture of the halo nucleus:
Since the halo neutron is very weakly bound,
it easily escapes from the binding of the projectile
and hardly changes its velocity,
while the $^{10}$Be core is decelerated by the Coulomb potential
from the target.
This leads to loss of the effective colliding energy of $^{10}$Be
because the escaped neutron carries about $1/11$
of the original projectile energy.
This accounts for obtained reduction of the fusion cross section.

In this article, we compare
the fusion process of neutron halo nuclei with
that of stable nuclei.
Different dynamical effects of a valence neutron are discussed and
we further elucidate a spectator picture of fusion dynamics of halo nuclei.

\section{TDWP method and applications}

We describe a projectile (P) as a two-body bound system of a core nucleus
(denoted as C) and a valence neutron (n).
In this work, the neutron is assumed to occupy a single-particle orbital of
$\ell=0$ ($s$-wave).
The projectile collides with a target (T) which is assumed to be
in the ground state all the time.
The time-dependent Schr\"odinger equation for the three-body model 
is given by
\begin{equation}
i\hbar \frac{\partial}{\partial t} \Psi({\bm R},{\bm r},t)
= 
\left\{ -\frac{\hbar^2}{2\mu} \nabla_{\bm R}^2
    -\frac{\hbar^2}{2m}   \nabla_{\bm r}^2 
    +V_{\rm nC}(r) +V_{\rm CT}(R_{\rm CT}) +V_{\rm nT}(r_{\rm nT}) \right\}
\Psi({\bm R},{\bm r},t),
\label{3BTDS}
\end{equation}
where the relative n-C coordinate is denoted as ${\bm r}$ and the
relative P-T coordinate as ${\bm R}$.
The reduced masses for the n-C and P-T motions are
$m$ and $\mu$, respectively.
The real potential, $V_{\rm nC}(r)$, produces a projectile and controls
binding energy of the neutron in the projectile.
The neutron-target potential, $V_{\rm nT}(r_{\rm nT})$, is also real.
This potential controls the orbital energies in the target and affects
properties of transfer reaction.
The core-target potential, $V_{\rm CT}(R_{\rm CT})$, consists of
nuclear and Coulomb potentials.
The nuclear part contains an imaginary part inside the Coulomb barrier
between the core and the target.
When the wave function has its amplitude inside the barrier,
a part of the flux will be lost.
This absorption describes the fusion of the core and target nuclei.
We define the fusion cross section in terms of the loss of flux, 
which is almost equivalent to the definition using the incoming boundary
condition inside the barrier.
Since we do not calculate the wave function of the fused system in this
framework,
we cannot distinguish the complete fusion (CF) process from the incomplete
one (ICF).
Thus, the fusion cross section we show in this article represents
the total fusion cross section (CF+ICF).

The wave function has six variables, ${\bm R}$ and ${\bm r}$.
We transform the equation in the laboratory frame to the one in the
body-fixed frame.
Then, the variables are three Euler angles ($\alpha,\beta,\gamma$),
an angle $\theta$ between $\bm R$ and $\bm r$,
and two radial variables, $R$ and $r$.
The wave function is expressed in a form
\begin{equation}
\Psi^{JM}(R,r,\alpha,\beta,\gamma,\theta;t) = 
\sum_\Omega^J \sum_\ell^{\ell_{\rm max}}
     \frac{u^J_{\Omega\ell}(R,r;t)}{Rr}
      G_{\Omega\ell}^{JM}(\alpha,\beta,\gamma,\theta) ,
\end{equation}
where the angle function $G^{JM}_{\Omega\ell}$ is expressed by
the $D$-function, $D^J_{\Omega M}({\bm\omega})$,
and the associated Legendre polynomials,
$P_\ell^\Omega(\cos\theta)$ \cite{Pack74,IO87}.
The wave-packet method is suited for the body-fixed representation
because it can take an advantage of the sparsity of
the Coriolis couplings.

In order to prepare an initial wave packet for the three-body system,
first, the wave function for projectile is constructed as an eigenstate of the
core-neutron Hamiltonian, $\phi_0(r)$.
Then, it is multiplied by
a localized Gaussian boosted toward the collision,
$u(R,r,t=0)=\phi_0(r)\exp(-(R-R_0)^2/2\lambda-iKR)$.
The parameters, $\lambda$ and $K$,
are chosen so as to cover an energy range of interest.
A single time evolution of the wave packet contains all the information
for this range of incident energy.
We perform calculations of the energy projection for the
wave packets before and after the collision,
to obtain the rate of flux loss as a function of incident energy.

The radial coordinates, $R$ and $r$, are discretized in mesh of
$0.2\sim 1$ fm.
The discrete variable representation is utilized in evaluation of the
differentiation.
The time evolution is achieved by the fourth-order Taylor expansion.
Details are found in our recent papers\cite{YUN03-P,NYIKU04-P,YIKUN04-P,IYNU06}.

\subsection{Case I: Stable projectile}

\begin{figure}
  \label{fig:prob_stab}
  \includegraphics[height=.3\textheight]{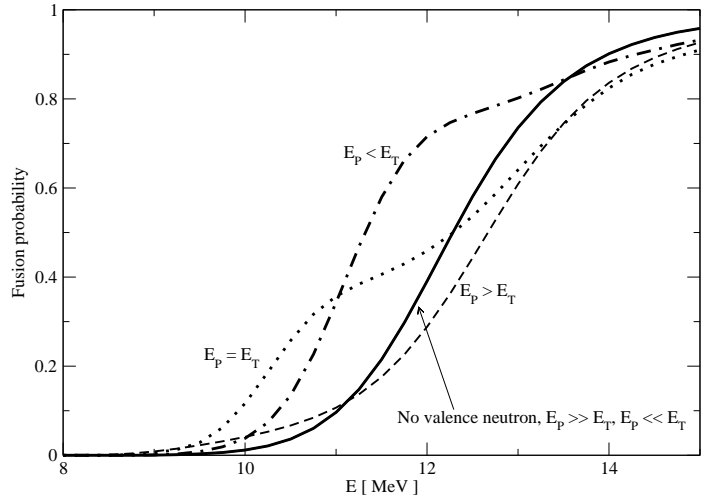}
  \caption{Fusion probability ($J=0$) as a function of incident energy for
  a tightly bound ``$^{11}$Be'' nucleus ($\epsilon_n=-3.5$ MeV) on $^{40}$Ca.
  Solid line indicates the probability for the projectile without
  the valence neutron.
  Dotted, Dashed, and Dash-dotted lines indicate fusion probabilities
  calculated with different
  $V_{\rm nT}$ potentials which give orbital energies in the target of
  $-3.5$, $-5.1$, $-2.0$ MeV, respectively.
  See text for details.}
\end{figure}

First, we investigate fusion reaction of a stable projectile.
For the projectile in which the valence neutron is rather tightly bound,
the fusion cross section is normally identical to that without the valence
neutron.
However, when the condition of energy matching is satisfied,
which means that energies of the neutron orbital in the projectile
and target are degenerate,
we may observe a substantial enhancement of sub-barrier fusion cross section.
This can be seen in Fig.~\ref{fig:prob_stab}.
The TDWP calculations are performed with a fictitious $^{11}$Be nucleus,
in which the valence neutron has a binding energy of
$3.5$ MeV at the $2s$ orbital.
The depth of $V_{\rm nC}(r)$ is increased to produce this situation.

\begin{figure}
  \label{fig:snap_stab}
  \includegraphics[height=.2\textwidth,angle=90]{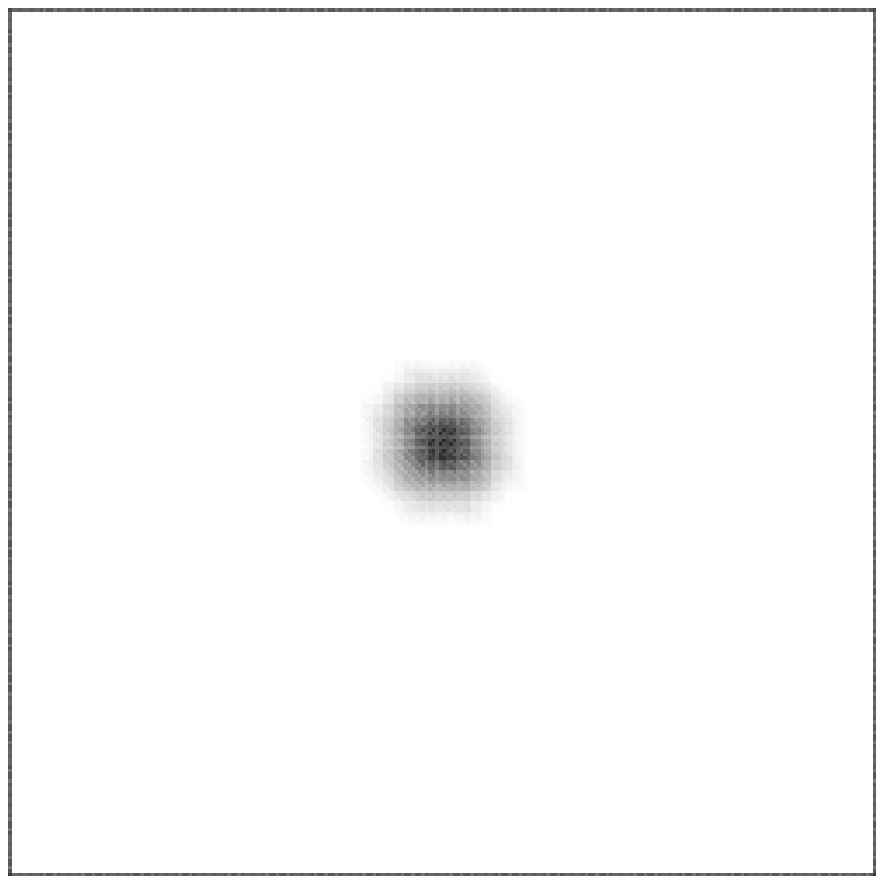}
  \includegraphics[height=.2\textwidth,angle=90]{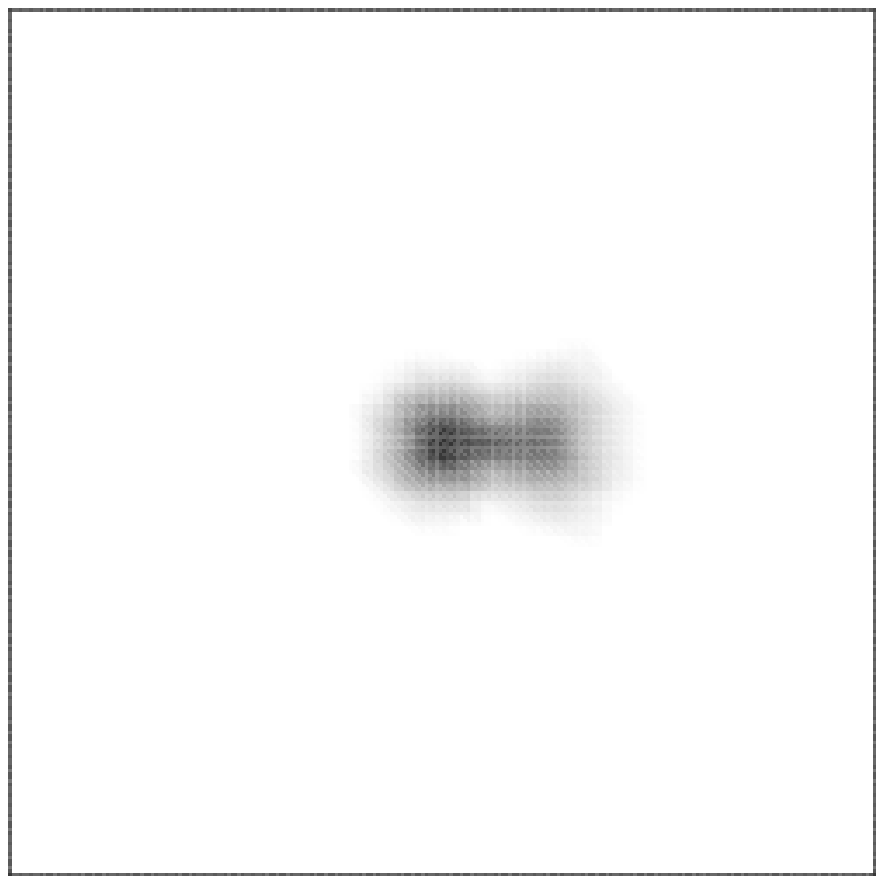}
  \includegraphics[height=.2\textwidth,angle=90]{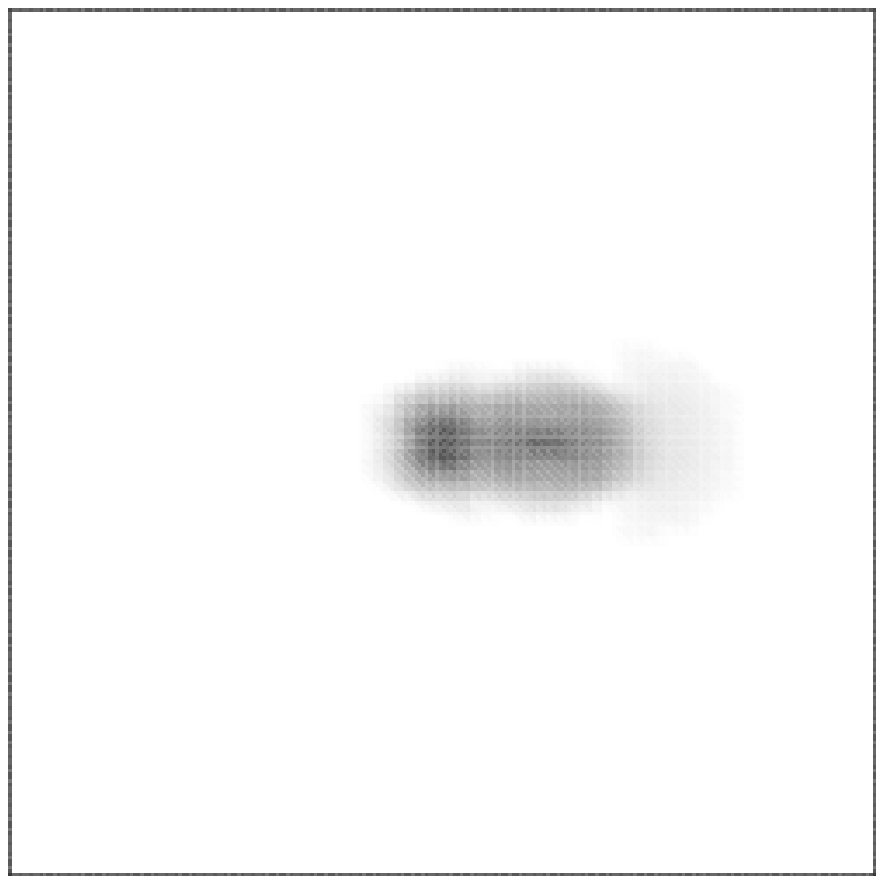}
  \includegraphics[height=.2\textwidth,angle=90]{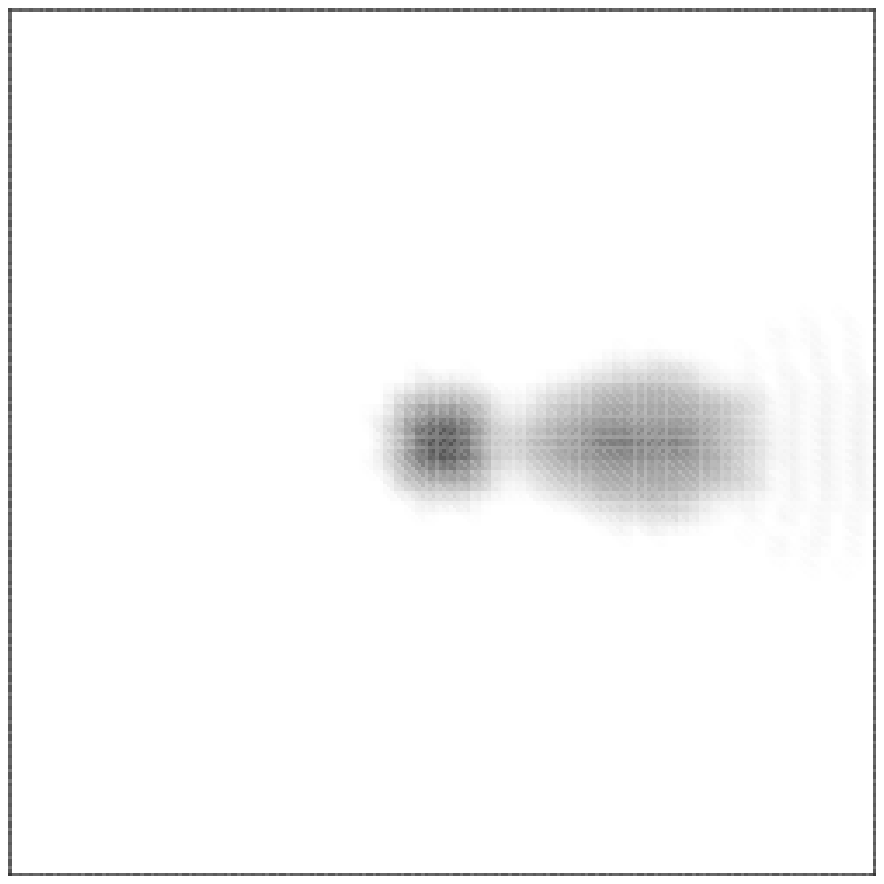}
  \includegraphics[height=.2\textwidth,angle=90]{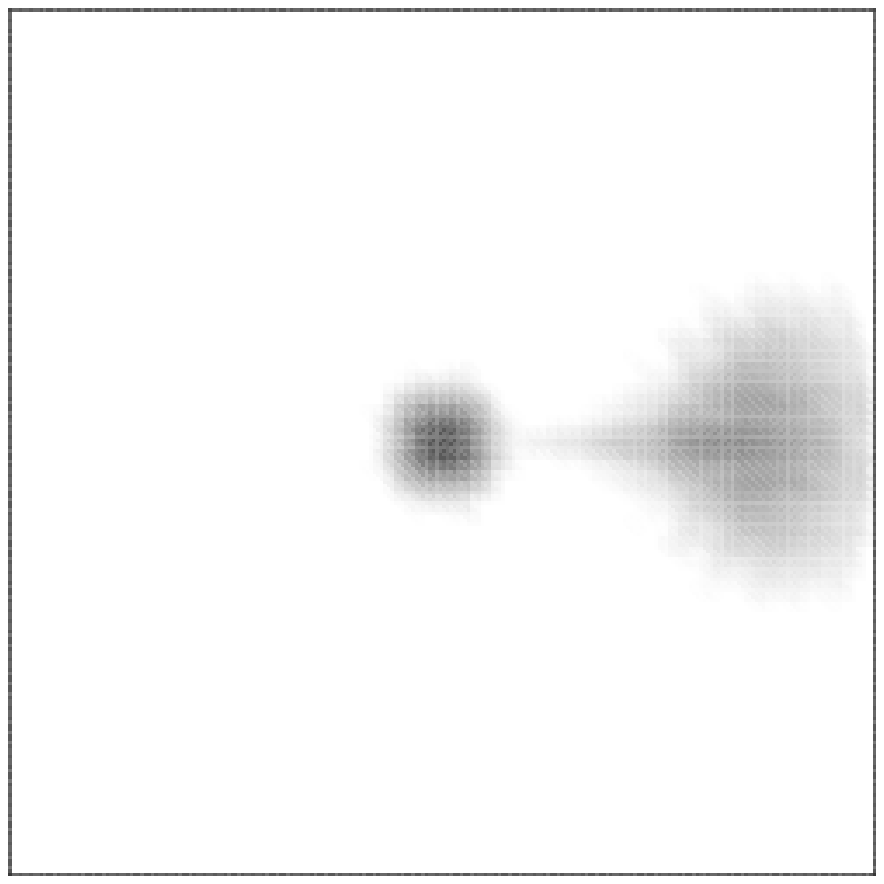}
  \caption{Time-dependent density distribution of the valence neutron,
  $\rho(r,\theta;t)$,
  corresponding to the dotted line in Fig.~\ref{fig:prob_stab}.
  The left-end panel is the initial distribution and the right-end
  shows the final distribution after the collision.
  The center of each panel, $r=0$, indicates the position of the core,
  and the right is the direction for the target, $\theta=0$.
  The target nucleus (wave packet) approaches toward the center
  (the first to second panels),
  then leaves away to the right (the third to fifth panels).
  }
\end{figure}

We vary the depth of the neutron-target potential, $V_{\rm nT}$, and
compare the fusion probabilities.
The results show that the fusion probability is correlated with the
neutron transfer probability and sensitive to $V_{\rm nT}$.
When the orbital energies in projectile and target are equal,
the fusion probability is enhanced
at energies below the barrier while it is suppressed above the barrier
(dotted line in Fig.~\ref{fig:prob_stab}).
This is caused by large neutron transfer from the projectile to the target.
Time evolution of the neutron density distribution during the
collision process is shown in Fig.~\ref{fig:snap_stab}.
Quantities, $\rho(r,\theta;t)=\int dR |\Psi^{00}(R,r,\theta;t)|^2$, are plotted
for the head-on collision of $J=0$.
Since the direction of the target always corresponds to $\theta=0$,
the target nucleus approaches from the right side of the panel, then
returns back to the right.
From the second to the fourth panels, we see that the neutron constitutes
a kind of molecular orbital extended over both projectile and target.
Since the energy matching is good in this case,
this results in the large neutron transfer at the final state.
The breakup component turns out to be small.

The sub-barrier fusion enhancement is even more pronounced when the
$V_{\rm nT}$ becomes slightly shallower than the exact matching case
(dash-dotted line in Fig.~\ref{fig:prob_stab}).
The opposite effect is observed if the $V_{\rm nT}$ is
made slightly deeper (dashed line).
This may suggest adiabatic dynamics of the valence neutron.
Namely, coupling between orbitals in projectile and target
either increase or decrease the kinetic energy of the projectile,
according to the relative position of orbital energies \cite{Yab97}.


\begin{figure}[t]
  \label{fig:snap_halo}
  \includegraphics[height=.2\textwidth,angle=90]{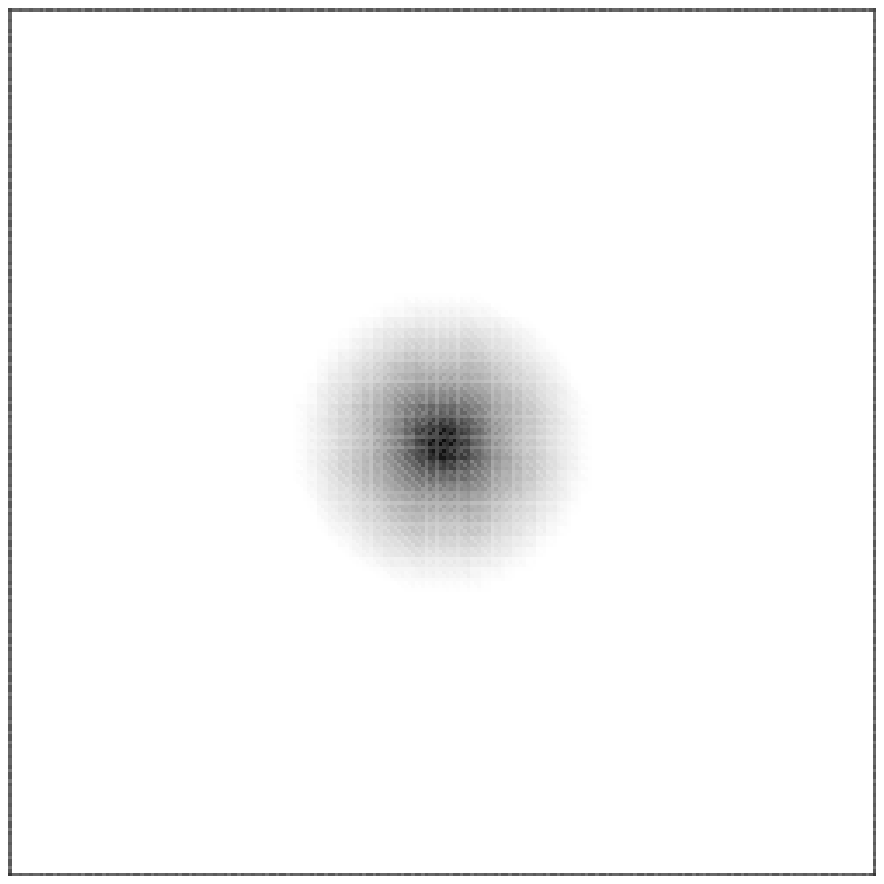}
  \includegraphics[height=.2\textwidth,angle=90]{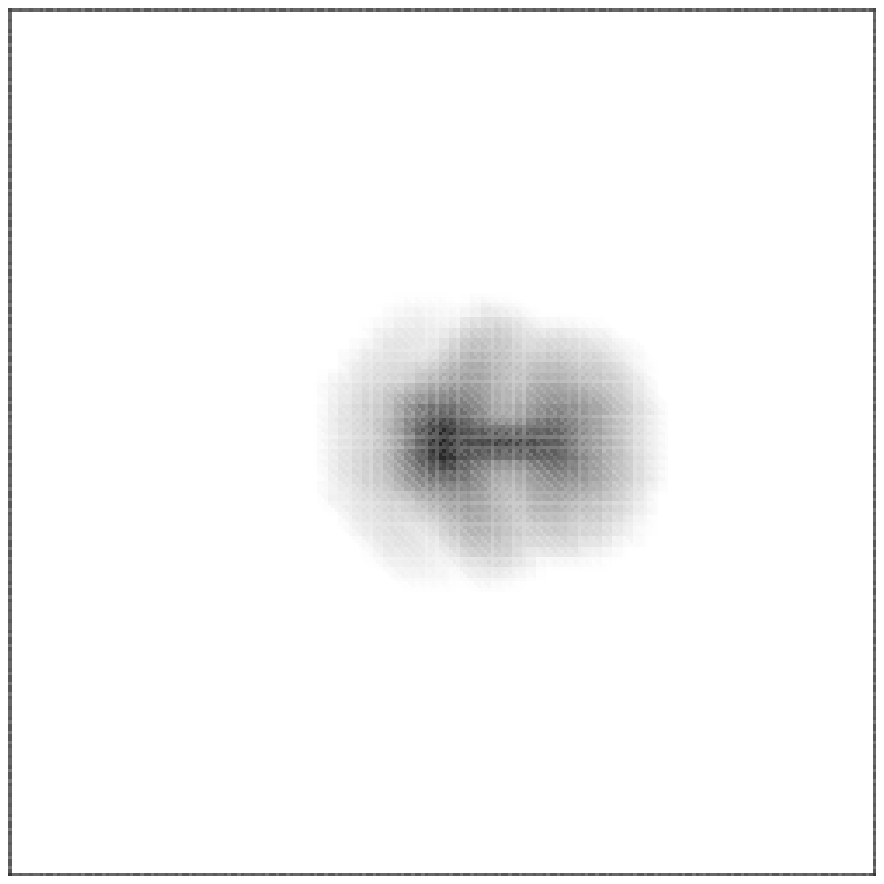}
  \includegraphics[height=.2\textwidth,angle=90]{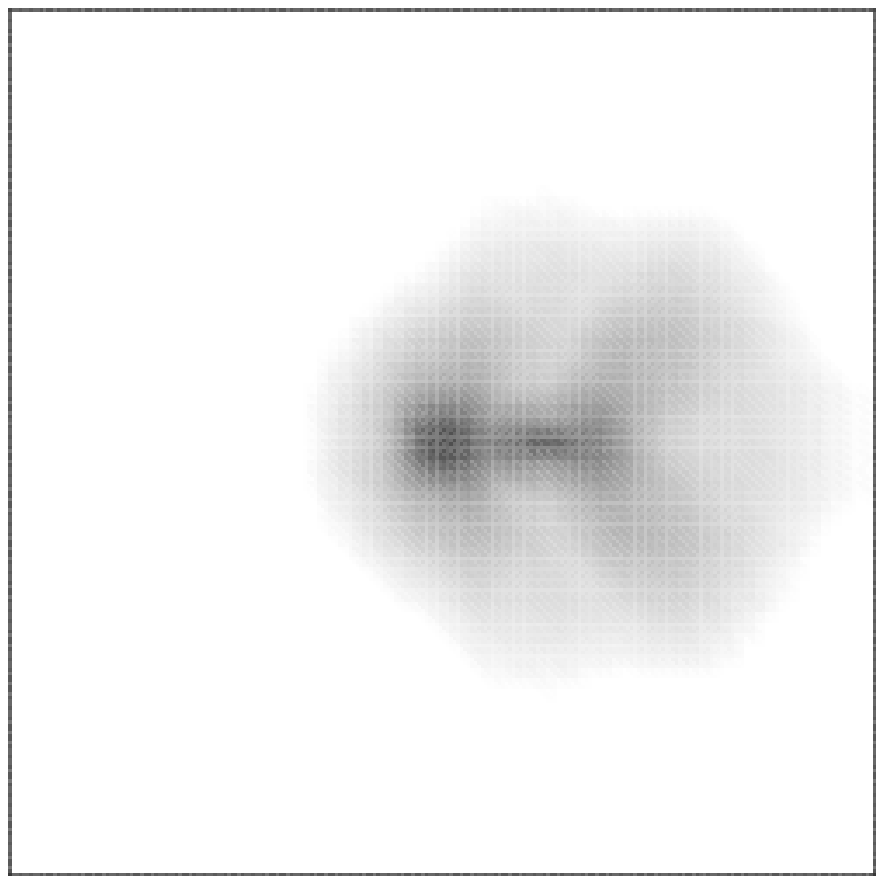}
  \includegraphics[height=.2\textwidth,angle=90]{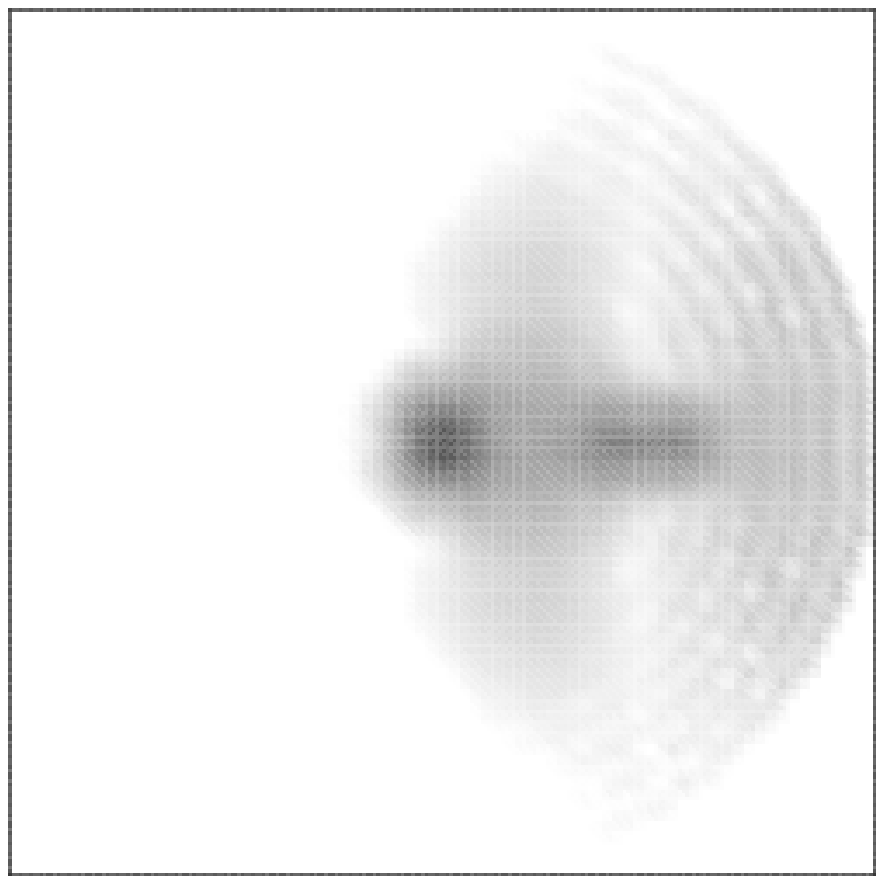}
  \includegraphics[height=.2\textwidth,angle=90]{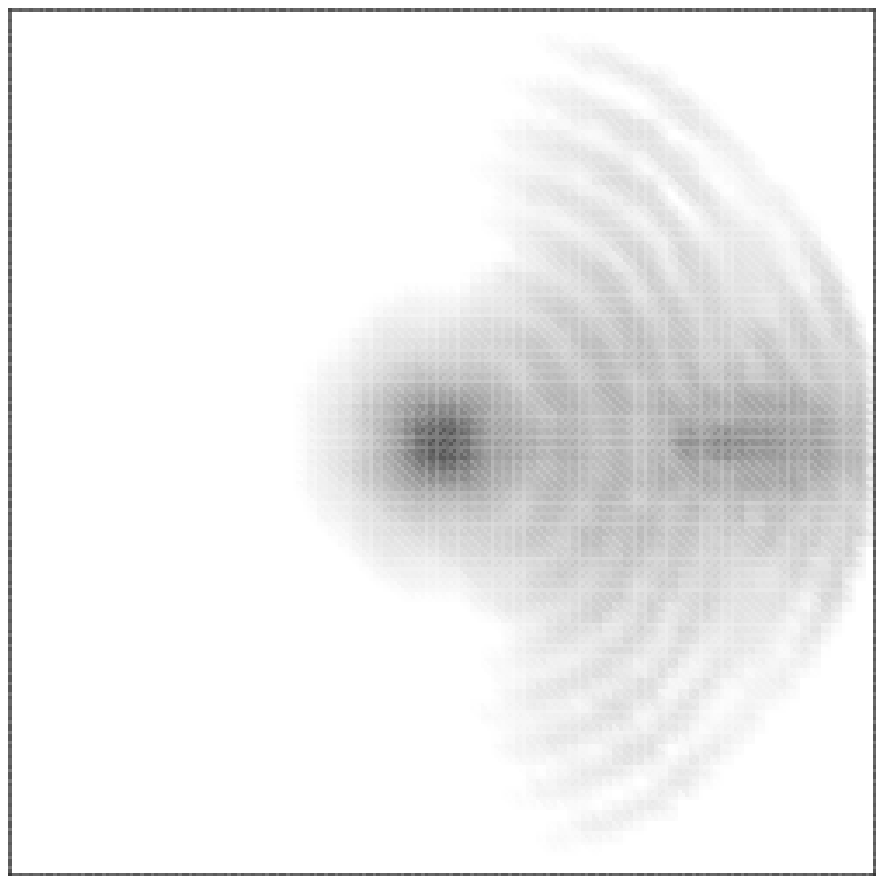}
  \caption{Time-dependent density distribution of the halo neutron,
  for the head-on collision, $J=0$, of halo $^{11}$Be on $^{209}$Bi.
  See caption of Fig.~\ref{fig:snap_stab}.
  }
\end{figure}
\begin{figure}[ht]
  \label{fig:prob_halo}
  \includegraphics[height=.35\textheight]{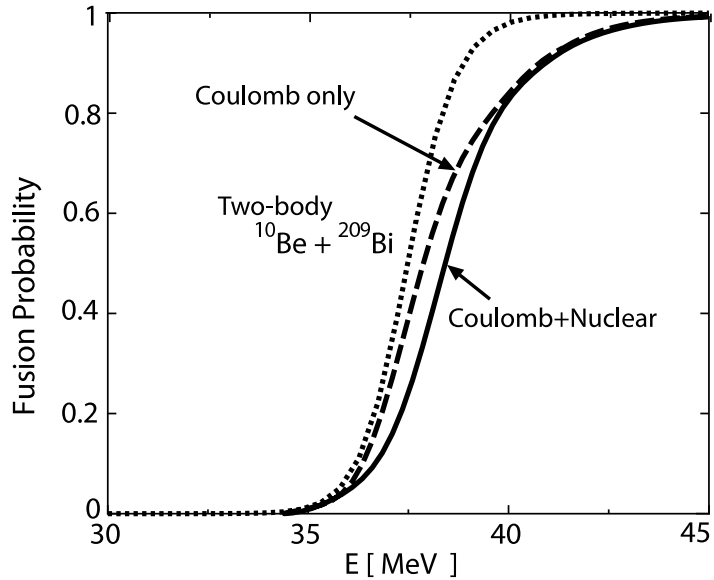}
  \caption{Fusion probability ($J=0$) as a function of incident energy for
  halo $^{11}$Be ($\epsilon_n=-0.6$ MeV) on $^{209}$Bi.
  Dotted line indicates the probability for the projectile without
  the valence neutron.
  Dashed and solid lines indicate fusion probabilities
  calculated with and without $V_{\rm nT}$ potentials,
  respectively.
  See text for details.}
\end{figure}
\subsection{Case II: Halo projectile}

In the previous section,
we have shown that, for the fusion reaction of stable projectiles,
the neutron transfer could play a role of ``glue'' at
sub-barrier energies,
if the energy matching is good between core and target orbitals.
Then, what about halo projectiles?
The neutron wave function is well extended out of the core nucleus,
producing a large r.m.s. radius.
Does this lead to even stronger enhancement of fusion probability?
As far as our three-body model is concerned, the answer is ``No''.
The spectator picture of halo neutron, that is given in the introduction
of the present article, can give an intuitive understanding of
dynamics of halo nuclei.

In Fig.~\ref{fig:snap_halo}, the time evolution of neutron density is shown
for a case of good energy matching between core and target orbitals.
The neutron energy is set at $-0.6$ MeV to describe halo $^{11}$Be.
Comparing this with Fig.~\ref{fig:snap_stab},
we may notice large breakup components of neutron in the halo case.
The breakup neutron is expanding in forward directions of
$|\theta|<90^\circ$.
This is consistent with the spectator role of the halo neutron.
Since the $V_{\rm nT}$ potential is chosen to satisfy the energy matching
between the core and target orbitals,
we may identify some transfer components in the panel
at the right end.
However, in contrast to the case of stable projectiles,
the transferred neutron does not play a role of ``glue'' at all.
In Fig.~\ref{fig:prob_halo}, the fusion probability is shown
as a function of energy.
As is seen in this figure, the neutron-target potential, $V_{\rm nT}$,
changes the fusion probability very little.
At all energies, the fusion probability of $^{11}$Be is suppressed
compared to that of $^{10}$Be.
This leads to a simple conclusion;
The halo neutron leads to a suppression of fusion cross section.

\section{Summary}

Effects of a valence neutron on fusion cross section have been studied with
the time-dependent wave-packet method in the three-body model.
For stable projectiles, the fusion cross section could be enhanced by
the presence of the valence neutron, if the orbital energies in the projectile
and target are almost equal.
However, there is no enhancement for the halo projectile.
The fusion cross section is suppressed by the halo neutron, which can be
explained by the spectator role of the halo neutron.
Comparison with experiments has been done in Ref.~\cite{IYNU06}.
At least, recent measurements indicate no enhancement of fusion
cross section for halo nuclei, $^{11}$Be on $^{209}$Bi \cite{Sig04}
and $^6$He on $^{238}$U \cite{Raa04}.
Our results are consistent with these data.




\begin{theacknowledgments}
 This work has been supported by the Grant-in-Aid for Scientific
 Research in Japan (Nos. 17540231 and 17740160).
 A part of the numerical calculations have been performed
 at SIPC, University of Tsukuba,
 at RCNP, Osaka University,
 and at YITP, Kyoto University.
\end{theacknowledgments}



\bibliographystyle{aipproc}   

\bibliography{fusion06,myself}

\IfFileExists{\jobname.bbl}{}
 {\typeout{}
  \typeout{******************************************}
  \typeout{** Please run "bibtex \jobname" to optain}
  \typeout{** the bibliography and then re-run LaTeX}
  \typeout{** twice to fix the references!}
  \typeout{******************************************}
  \typeout{}
 }

\end{document}